\documentclass[a4paper,11pt]{article}
\usepackage{jheppub} 
\usepackage[utf8]{inputenc}
\usepackage{hyperref}
\usepackage{verbatim,graphics,graphicx,color,slashed,amsmath}
\usepackage[T1]{fontenc}
\usepackage{subfigure}
\usepackage[table,svgnames]{xcolor}
\usepackage{color}
\usepackage{amsfonts}
\usepackage{amssymb}
\usepackage{bbding}
\usepackage{appendix}
\usepackage{mhchem}
\usepackage{stmaryrd}
\usepackage{blkarray}
\usepackage[export]{adjustbox}
\graphicspath{ {./images/} }
\usepackage{bbold}
\usepackage{CJKutf8}
\usepackage{bm}
\usepackage{orcidlink}
\usepackage{dcolumn}
\usepackage{multirow}
\usepackage{capt-of}
\usepackage{mathrsfs}
\usepackage{enumitem}
\usepackage{comment}
\usepackage{float}
\usepackage{array}
\usepackage{dsfont}
\usepackage{physics}
\usepackage[section]{placeins}
\usepackage{epsfig,psfrag,rotating,soul}
\usepackage{rotfloat}
\usepackage{empheq}  
\usepackage[colorlinks=true,linkcolor=blue,urlcolor=blue,citecolor=blue]{hyperref}
\numberwithin{equation}{section}
\newcommand{\be}{\begin{equation}}
\newcommand{\ee}{\end{equation}}
\newcommand{\bea}{\begin{eqnarray}}
\newcommand{\eea}{\end{eqnarray}}
\newcommand{\ba}{\begin{aligned}}
\newcommand{\ea}{\end{aligned}}

\allowdisplaybreaks

\title{\boldmath How a minimal length scale  modifies thermodynamics of RN AdS Black Holes}

\author[a,b]{Suman Kumar Panja \orcidlink{0000-0001-9258-9825}}
\author[a,b]{\,and \, Yu Shi}
\affiliation[a]{Wilczek Quantum Center, Shanghai Institute for Advanced Studies, Shanghai 201315, China }
\affiliation[b]{University of Science and Technology of China, Hefei 230026, China}

\emailAdd{sumanpanja@ustc.edu.cn}
\emailAdd{yu\_shi@ustc.edu.cn}

\abstract{We investigate the thermodynamic modifications of the Reissner-Nordstr\"om anti-de Sitter (RN AdS) black hole induced by a minimal length scale, which naturally emerges in $\kappa$-deformed space-time. By constructing the modified metric via phase-space commutation relations, we derive the deformed Hawking temperature, entropy, and enthalpy. We analyze the thermal stability through the heat capacity and study the $P$-$V$ criticality, revealing that the black hole undergoes a small-to-large phase transition analogous to the Van der Waals system, albeit with a critical ratio slightly lowered by non-commutativity. Furthermore, we examine the Joule–Thomson expansion and find that the non-commutative (NC) parameter expands the cooling region in the temperature-pressure plane. Our results demonstrate that while the overall thermodynamic analogy with the Van der Waals fluid persists, the minimal length scale systematically deform the coexistence region and inversion curves, offering potential observational signatures for quantum gravity.}

\begin{document}
\maketitle
\flushbottom

\section{Introduction}\label{sec:1}

Black holes represent the most extreme gravitational environments in the universe. The seminal works of Bekenstein and Hawking established that black holes are thermodynamic systems, possessing entropy proportional to their horizon area and radiating thermally \cite{Bek-1, Hawking-1}. This profound connection serves as a crucial bridge among general relativity, quantum mechanics, and statistical physics \cite{S-5}.

The thermodynamics of anti-de Sitter (AdS) black holes has attracted particular attention due to the AdS/CFT correspondence. Hawking and Page first demonstrated a phase transition between the Schwarzschild-AdS black hole and thermal AdS space \cite{Page}. A major breakthrough came with the recognition that charged AdS black holes exhibit $P$-$V$ criticality completely analogous to the Van der Waals liquid-gas system, where the cosmological constant is interpreted as thermodynamic pressure \cite{Charged-AdS, Mann}. This analogy has been extended to rotating, higher-dimensional, and Born-Infeld black holes \cite{Mann-2}, and more recently to the Joule–Thomson expansion \cite{C-AdS} and Landau-functional descriptions of phase transitions \cite{C-AdS-1}.

At Planckian scales, space-time is expected to be discrete, prompting modifications to classical black hole thermodynamics. Non-commutative geometry provides a natural framework to incorporate the minimal length scale while preserving certain symmetries \cite{connes, Glikman, dop}. Among various NC space-times, the $\kappa$-deformed Minkowski space-time, characterized by the Lie-algebraic commutation relations 
\be
[\hat{x}^i,\hat{x}^j]=0,~~~[\hat{x}^0, \hat{x}^i]=ia\hat{x}^i,~~~a=\frac{1}{\kappa}, \label{com2}
\ee 
has received growing attention because it accommodates a fundamental length parameter $a$ without entirely breaking Lorentz invariance at the Hopf-algebra level \cite{kappa1, mel1}. Recent studies have explored $\kappa$-deformed effects on Hawking radiation \cite{Moyal-1, zuhair1}, Schwarzschild-AdS thermodynamics \cite{NC-BH, vishnu}, and even neutron-star properties \cite{neutronstarcosmo}. However, a systematic investigation of the $\kappa$-deformed RN AdS black hole, particularly its Joule–Thomson expansion and the full phase structure in the extended phase space, remains absent.

In this work, we fill this gap by examining how the minimal length scale modifies the thermodynamics, stability, critical phenomena, and Joule–Thomson expansion of the RN AdS black hole in $\kappa$-deformed space-time. We construct the modified metric, compute all relevant thermodynamic quantities, and systematically compare the deformed behavior with both the commutative RN AdS case and the Van der Waals fluid. Our study reveals that while the NC deformation preserves the qualitative Van der Waals analogy, it quantitatively shifts critical points and inversion curves, enlarging the cooling region in the throttling process.

The organization of this paper is as follows. The next section briefly reviews $\kappa$-deformed space-time and the construction of $\kappa$-deformed RN AdS black hole metric. In section \ref{sec:3}, we calculate the NC modifications to the thermodynamic quantities and examine the thermal behavior of the $\kappa$-deformed RN AdS black hole. Further, we investigate the black hole's stability by analyzing the deformed heat capacity. Section \ref{sec:4} studies the critical phenomena and phase transition through examination of $P-V$ criticality and variation of Gibbs free energy with temperature. In section \ref{sec:5}, the investigation of Joule–Thomson expansion and the cooling-heating region of the $\kappa$-deformed RN AdS black hole is given. Concluding remarks are given in section \ref{sec:6}.

\section{$\kappa$-deformed RN AdS black hole metric}\label{sec:2}

In this section, we first give a summary of the $\kappa$-deformed space-time. Then we construct the $\kappa$-deformed metric for the RN AdS black hole. $\kappa$-deformed space-time is a Lie-algebraic type NC space-time. The coordinates of this NC space-time adhere to the commutation relations given in Eq.(\ref{com2}). The NC coordinates ($\hat{x}_{\mu}$) of $\kappa$-deformed space-time can be expressed in terms of commutative coordinates and their corresponding derivatives as \cite{mel1}
\be \label{ksp-2}
 \hat{x}_0=x_0\psi(ap^{0})+iax_j\partial_j\gamma(ap^{0}), ~\text{and}~~ \hat{x}_i=x_i\varphi(ap^{0}),
\ee
where, $\psi$, $\gamma$, and $\varphi$ adhere to $\psi(0)=1,~\varphi(0)=1$. Here, $p^{0}$ is the energy associated with the commutative space-time, and from this point, we redefine $p^{0}$ as $\varepsilon$. Employing the above equation in Eq.(\ref{com2}) one finds $\frac{\varphi'(a\varepsilon)}{\varphi(a\varepsilon)}\psi(a\varepsilon)=\gamma(a\varepsilon)-1$ with $\varphi^{\prime}=\frac{d\varphi}{d(a\varepsilon)}$. Possible realizations for $\psi(a\varepsilon)$ are $\psi(a\varepsilon)=1$ and $\psi(a\varepsilon)=1+2a\varepsilon$ \cite{mel1}. In this study, we choose $\psi(a\varepsilon)=1$. The allowed choices of $\varphi$ include $e^{-a\varepsilon}, e^{-\frac{a\varepsilon}{2}}, 1, \frac{a\varepsilon}{e^{a\varepsilon}-1}$, etc. \cite{mel1}. For our investigation, we choose $\varphi(a\varepsilon)=e^{-a\varepsilon}$. This particular choice corresponds to a specific ordering of the NC algebra and is commonly adopted in the literature; the resulting deformation parameter $a\varepsilon$ acts as an effective dimensionless parameter, with $\varepsilon$ related to the black hole mass.

Next, to obtain the $\kappa$-deformed metric for the RN AdS black hole, we start with the general form of the commutation relation for positions and momenta in $\kappa$-deformed space-time \cite{kappa-geod}
\be \label{N1}
 [\hat{x}_{\mu},\hat{P}_{\nu}]=i\hat{g}_{\mu\nu}, 
\ee
where $\hat{g}_{\mu\nu}(\hat{x}^{\alpha})$ represents $\kappa$-deformed metric. The phase-space coordinates of the $\kappa$-deformed space-time are written as \cite{kappa-geod},
\be \label{N2}
 \hat{x}_{\mu}=x_{\alpha}\varphi^{\alpha}_{\mu}, \,\hat{P}_{\mu}=g_{\alpha\beta}(\hat{y})p^{\beta}\varphi^{\alpha}_{\mu},
\ee
where $\hat{P}_{\mu}$ and $p_{\mu}$ are the canonical conjugate momenta corresponding to the NC coordinates $\hat{x}_{\mu}$ and  the commutative coordinate $x_{\mu}$, respectively. 
In the above equation we have introduced another set of $\kappa$-deformed space-time coordinates $\hat{y}_{\mu}$, which is assumed to adhere $\hat{x}_{\mu}$, i.e., $[\hat{y}_{\mu},\hat{x}_{\nu}]=0$ and can be written as  $\hat{y}_0=x_0-ax_jp^j,~~\hat{y}_i=x_i$ (for details see \cite{zuhair1,kappa-geod}). Using Eq.(\ref{N2}) in Eq.(\ref{com2}) we find a specific realization for $\varphi_{\mu}^{\alpha}$ to be
\be \label{N3}
 \varphi _0^0=1, \, \varphi _i^0=0, \, \varphi_0^i=0, \, \varphi _j^i=\delta _j^i e^{-a\varepsilon}. 
\ee
Further substituting the Eq.(\ref{N2}) in Eq.(\ref{N1}), one find the $\kappa$-deformed metric as \cite{zuhair1}
\begin{equation}\label{N7}
 [\hat{x}_{\mu},\hat{P}_{\nu}] \equiv i\hat{g}_{\mu\nu}=ig_{\alpha\beta}(\hat{y})\Big(p^{\beta}\frac{\partial \varphi^{\alpha}_{\nu}}{\partial p^{\sigma}}\varphi_{\mu}^{\sigma}+\varphi_{\mu}^{\alpha}\varphi_{\nu}^{\beta}\Big). \end{equation}
Here, $g_{\mu \nu}(\hat{y})$ is found by taking the commutative metric and replacing its coordinates with the NC coordinates $\hat{y}_{\mu}$. Next, we substitute Eq.(\ref{N3}) in the above equation and obtain the generic form of the $\kappa$-deformed line element for spherically symmetric space-time to be \cite{zuhair1} 
\be \label{N12}
 d\hat{s}^2=g_{00}dx^0dx^0+g_{ij}e^{-4a\varepsilon}dx^idx^j. 
\ee
In this study, we investigate thermodynamic properties of RN AdS black holes (characterized by their mass $M$ and charge $Q$) in the background of $\kappa$-deformed space-time. The solution of the four-dimensional RN AdS black hole is defined by the line element \cite{Charged-AdS},
\be
ds^2 = -f(r)dt^2 + \frac{dr^2}{f(r)} + r^2(d\theta^2 + \sin^2\theta\,d\phi^2), \label{N11a}
\ee
where $f(r) = 1 - \frac{2M}{r} + \frac{Q^2}{r^2} + \frac{r^2}{L^2}$ and $L$ is the AdS radius, which is related to the negative cosmological constant as $\Lambda = - \frac{3}{L^2}$. Using Eq.(\ref{N11a}) in Eq.(\ref{N12}), we find the $\kappa$-deformed line element for RN AdS black hole as
\be
d\tilde{s}^2 = -f(r)dt^2 + e^{-4a\varepsilon}\left[\frac{dr^2}{f(r)} + r^2(d\theta^2 + \sin^2\theta\,d\phi^2)\right]. \label{N12a}
\ee
In the commutative limit, i.e., $a \rightarrow 0$, the above expression of the deformed metric reduces to Eq.(\ref{N11a}).

\section{Modified thermodynamics of black hole and stability analysis}\label{sec:3}

\begin{figure}[h!]
    \centering
    \includegraphics[width=0.48\textwidth]{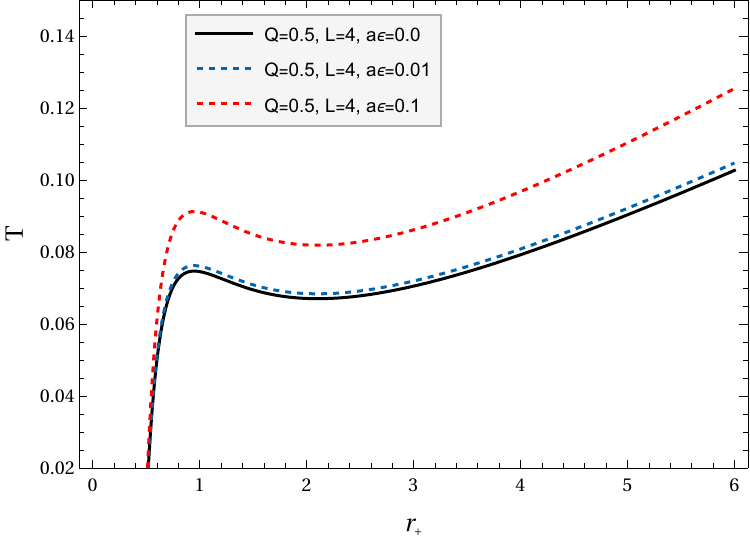}
    \includegraphics[width=0.48\textwidth]{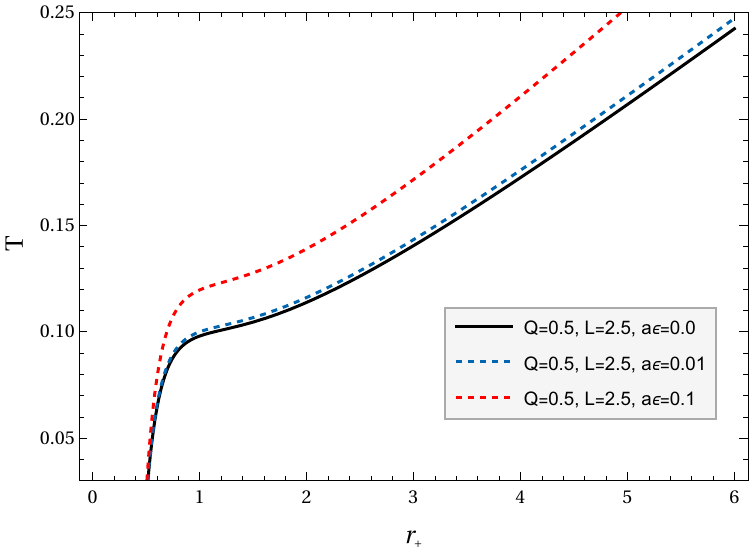}
   \caption{Effect of non-commutativity on the variation of the Hawking temperature ($T$) as a function of horizon radius ($r_{+}$) for a fixed charge $Q=0.5$, considering two different values of the Cosmological constant (i.e., $L=4$ and $L=2.5$).}
    \label{fig-2}
\end{figure}

\begin{figure}[h!]
    \centering
    \includegraphics[width=0.48\textwidth]{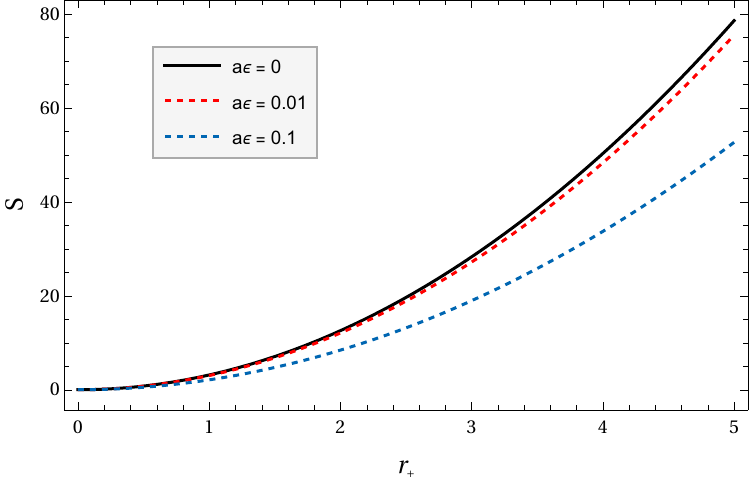}
  \caption{Variation of entropy ($S$) as a function of horizon radius ($r_{+}$) considering different values of $a\varepsilon$.}
    \label{fig-1}
\end{figure}

In this section, to understand the thermal behavior and explore the stability of the $\kappa$-deformed RN AdS black hole, we calculate and analyze thermodynamic quantities, including mass, temperature, entropy, enthalpy, and heat capacity. We commence with $\kappa$-deformed RN AdS black hole metric given in Eq.(\ref{N12a}) and consider that the event horizon radius \( r_+ \) satisfies \( f(r_+) = 0 \) to find the mass-horizon relation as 
\be
M = \frac{r_+}{2}\left(1 + \frac{r_+^2}{L^2} + \frac{Q^2}{r_+^2} \right). \label{N13}
\ee
Note here that the mass-horizon relation does not get modified due to the non-commutativity of the space-time. To calculate Hawking temperature, using Eq.(\ref{N12a}) we first find the surface gravity, $\kappa = \sqrt{-\tilde{g}^{rr}\tilde{g}^{tt}(\frac{1}{2}\partial_{r}\tilde{g}_{tt})^2} \Bigg |_{r=r_{+}} = \frac{e^{2a\varepsilon}}{2} f'(r_+),$ where $f'(r_+) = \frac{1}{r_+}\left(1 + \frac{3r_+^2}{L^2}- \frac{Q^2}{r_+^2} \right)$. Now using this expression of the surface gravity, we obtain the deformed Hawking temperature as
\be
T = \frac{\kappa}{2\pi} = \frac{e^{2a\varepsilon}}{4\pi r_+}\left(1  + \frac{3r_+^2}{L^2} - \frac{Q^2}{r_+^2}\right). \label{N17}
\ee
In the above equation, the absolute value of the Hawking temperature gets modified due to the space-time non-commutativity of space-time. The similar NC modification to the Hawking temperature has been obtained using the method of Bogoliubov coefficients by following the procedure discussed in \cite{zuhair1}. In Fig-\ref{fig-2}, we investigate the influence of space-time non-commutativity on the variation of Hawking temperature as a function of horizon radius for two different values of Cosmological constant ($\Lambda =-\frac{3}{L^2}$) with fixed charge. We observe that with an increase of the NC parameter (i.e., $a\varepsilon =0$ to $a\varepsilon=0.1$), the absolute value of the Hawking temperature increases, implying faster evaporation and a shorter lifetime of the deformed black hole compared to the standard RN AdS black hole.
Next to obtain the entropy using Bekenstein-Hawking area law ($S=\frac{\mathcal{A}}{4}$) \cite{Bek-1,Hawking-1}, we first obtain the surface area of the $\kappa$-deformed RN AdS black hole event horizon as
\be
\mathcal{A} = \int_0^{2\pi} d\phi \int_0^{\pi} \sqrt{\tilde{g}_{\theta\theta}\tilde{g}_{\phi\phi}}\, d\theta = e^{-4a\varepsilon} 4\pi r_+^2, \label{N14}
\ee
where $\tilde{g}_{\theta\theta} = e^{-4a\varepsilon} r^2$ and  $\tilde{g}_{\phi\phi} = e^{-4a\varepsilon} r^2 \sin^2\theta$ coming from Eq.(\ref{N12a}). Thus, using the above, we find the modified entropy to be
\be
S = \pi e^{-4a\varepsilon} r_+^2 . \label{N15}
\ee
Note that entropy gets modified due to the non-commutativity of the space-time. We analyze the variation of the entropy as a function of horizon radius in the presence of $\kappa$-deformation in Fig-\ref{fig-1}. We notice that the absolute value of the entropy decreases as the value of the NC parameter ($a\varepsilon$) increases, indicating a reduction in the effective number of microstates available to the $\kappa$-deformed RN AdS black hole.
The pressure associated with the black hole thermodynamics is related to the cosmological constant as
\be
P = -\frac{\Lambda}{8\pi} = \frac{3}{8\pi L^2}. \label{N18}
\ee
Next, using the extended first law of black hole thermodynamics, $ dH = TdS + VdP + \Phi dQ $ (where $\Phi$ is the electric potential measured at infinity with respect to the horizon) and using Eq.(\ref{N15}), we obtain the enthalpy of the system to be
\be
H = \frac{e^{-2a\varepsilon} r_+}{2}\left(1 + \frac{Q^2}{r_+^2} + \frac{r_+^2}{L^2}\right),\label{N19}
\ee
which adhere to $ S=\int^{r_{+}}_{0} \frac{1}{T}(\frac{\partial H}{\partial r})_{Q,P}$. Using the above, the thermodynamic volume of the $\kappa$-deformed RN AdS black hole and electric potential can be obtained as
\be
V = \left(\frac{\partial H}{\partial P}\right)_{S,Q} = e^{-2a\varepsilon} \frac{4\pi}{3} r_+^3, \label{N20a}
\ee
and
\be
\Phi = \left(\frac{\partial H}{\partial Q}\right)_{S,P} = \frac{e^{-2a\varepsilon} Q}{r_+}, \label{N20}
\ee
respectively. With these thermodynamic variables obtained above, the solution adheres to the first law of black hole thermodynamics in an extended phase space $dH=TdS + VdP +\Phi dQ$. Further using Eq.(\ref{N15})-Eq.(\ref{N20}), it is very straightforward to show that the Smarr relation $H=2TS-2PV + \Phi Q$ also holds, and the same can be derived using the scaling argument \cite{smarr}. It is important to note that all thermodynamic quantities are getting corrections due to the non-commutativity of the space-time. In the limit $a \rightarrow 0$, all these modified thermodynamics quantities reduce to commutative results obtained in \cite{C-AdS}. Apart from the Hawking temperature in Eq.(\ref{N17}), the absolute value of all other thermodynamic quantities ($S,V,H$) decreases due to space-time non-commutativity for considering a positive value of the deformation parameter $a>0$. Correction due to non-commutativity enhances the absolute value of the Hawking temperature. But a negative value of the deformation parameter (i.e., $a<0$) will increase magnitude of entropy and decrease the absolute value of the Hawking temperature. This can be interpreted as more microstates and slower evaporation for the $\kappa$-deformed black hole. 

\begin{figure}[h!]
    \centering
    \includegraphics[width=0.48\textwidth]{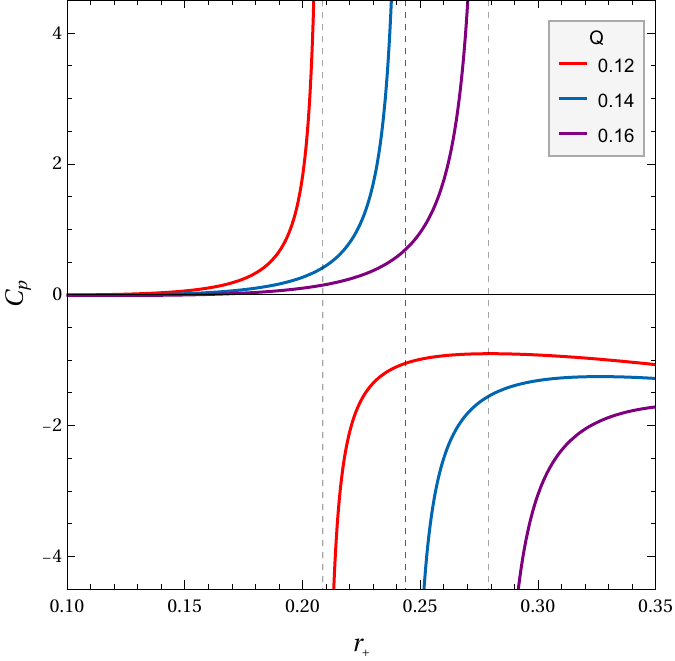}
    \includegraphics[width=0.47\textwidth]{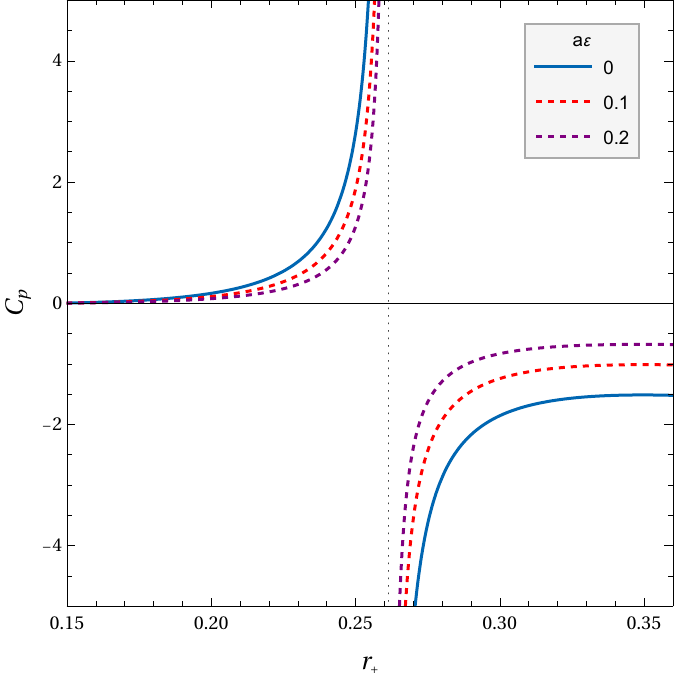}
    \caption{Variation of $\kappa$-deformed heat capacity ($C_{P}$) as a function of horizon radius ($r_{+}$) considering (a) different values of charge $Q$ (for fixed $a \varepsilon = 0.01$ and $L=4$) and (b) different values of $a \varepsilon$ (for fixed charge $Q=0.15$ and $L=4$). The vertical dashed lines mark the divergence points corresponding to second-order phase transitions.}
    \label{fig-4}
\end{figure}

Next to understand the thermodynamic stability of the black hole, employing Eq.(\ref{N17}) and Eq.(\ref{N15}), we obtain the deformed heat capacity at constant pressure ($C_{P}=T\Big(\frac{\partial S}{\partial T}\Big)_{P}$) as
\be
C_P = e^{-4a\varepsilon}2\pi r_{+}^2 \frac{1 - \frac{Q^2}{r_+^2} + \frac{3r_+^2}{L^2}}{\frac{3r_+^2}{L^2} - 1 + \frac{3Q^2}{r_+^2}}. \label{N21}
\ee
The above expression of modified heat capacity will reduce to the commutative result \cite{C-AdS} in the limit $a \rightarrow 0$. To understand the thermodynamic stability of the system, we investigate the variation in heat capacity for two cases, as shown in Fig-\ref{fig-4}.
In the first case, we vary heat capacity with $r_{+}$ for $a\varepsilon=0.01$ and different values of $Q$. For each curve, the region left of the dotted line represents a stable state (where $C_{p}<0$), while the area to the right shows an unstable state (where $C_{p}>0$). The dotted line marks the exact horizon radius $r_{+}$ where the black hole has a second-order phase change. As the charge increases, we observe a shift in the curves. This shift indicates that both the singularity and the stable to unstable transition occur at a larger horizon radius, similar to that observed in the commutative case. Similar analysis can be made for other values of $a\varepsilon$. In the second case, we examine how the heat capacity changes as $a\varepsilon$ is varied with the charge fixed. For very large horizon radius, the curves tend to overlap, indicating that NC effects are significant for smaller black holes and become less important as the black hole gets bigger. In this case, the $a\varepsilon$ dependent correction only changes the magnitude and slope of the curves, but does not change the roots of the denominator in Eq.(\ref{N21}). As a result, both the solid blue line (commutative case) and the dashed red lines (NC cases) have the same vertical asymptote at $r_+$.

\section{Modified P-V Criticality and Phase transitions}\label{sec:4}

In this section, we investigate critical phenomena and phase structure of the system using the thermodynamic equation of state. This method leads to a complete thermodynamic description of the system, similar to that for usual thermodynamic systems, such as Van der Waals fluids. For this, we start with the equation of state (coming from Eq.(\ref{N17})) that adheres to the P-V criticality of our research model
\be
P = \frac{1}{8\pi r_{+}^{2}} \left( e^{-2a\varepsilon}4\pi r_{+}T + \frac{Q^2}{r_+^2} -1 \right).\label{N22}
\ee
In the limit, $a \rightarrow 0$, the above expression of the $\kappa$-deformed pressure will reduce to the result obtained in \cite{C-AdS,Mann}. Next, we employ the prescription discussed in \cite{Mann} of translating the `geometric' equation of state (given in the above equation) to a physical one, and after calculational simplification, we obtain the equation of state as 
\be 
P = \frac{e^{-2a\varepsilon}T}{v}-\frac{1}{2\pi v^2} + \frac{2Q^2}{\pi v^4}, \label{N23}
\ee
where $v=2l_{p}^2 r_{+}$ ( the Planck length reads $l_{P}^2 = \frac{\hbar G}{c^3}$) is the specific volume, which has been introduced from comparison with the Van der Waals equation (see details in \cite{Mann}). Next to find the critical points for the deformed system, we analyze the equation of the state and solve for the conditions where first and second derivatives of pressure with respect of specific volume vanish, i.e., $\frac{\partial P}{\partial v}=0$ and $\frac{\partial^2 P}{\partial v^2}=0$, respectively. Thus, we find the critical point to be 
\be 
T_{c}= \frac{e^{2a\varepsilon}}{3\sqrt{6} \pi Q},~~ v_{c}=2\sqrt{6}Q, ~~  \text{and} ~~ P_{c}=\frac{1}{96 \pi Q^2}. \label{N24}
\ee
Using the above critical values, we obtain the critical ratio valid up to the first order in the deformation parameter $a$ as
\be 
\frac{P_{c}v_{c}}{T_{c}}=(1-2a\varepsilon)\frac{3}{8}. \label{N25}
\ee
The critical ratio is smaller than that of the Van der Waals fluid system as a result of the non-commutativity of space-time, similar to that reported in \cite{NC-BH,vishnu}. The equation above reduces to the critical ratio for the Van der Waals fluid in the limit $a \rightarrow 0$ \cite{Mann}. This analysis of criticality demonstrates that the black hole exhibits a small black hole to large black hole phase transition analogous to the Van der Waals system.
\begin{figure}[h!]
    \centering
    \includegraphics[width=0.429\textwidth]{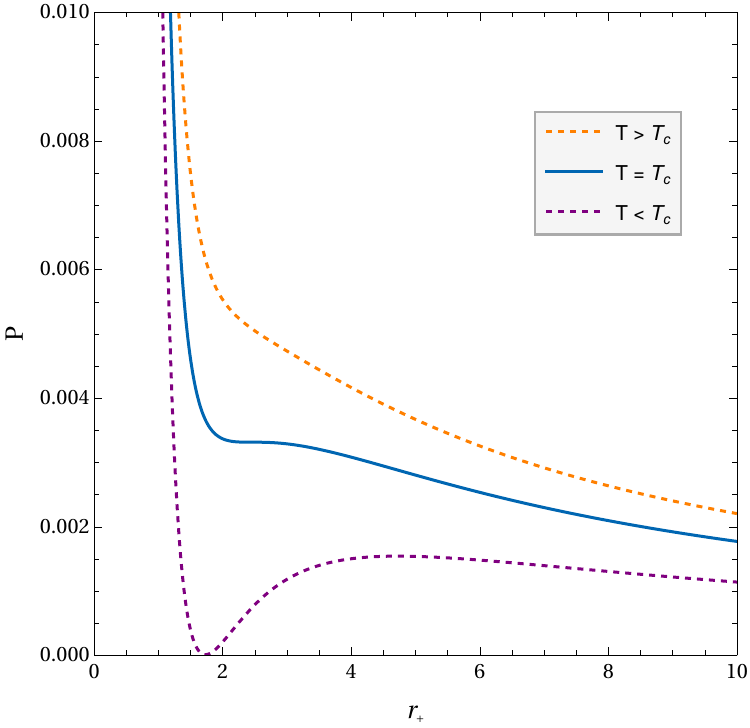}
    \includegraphics[width=0.422\textwidth]{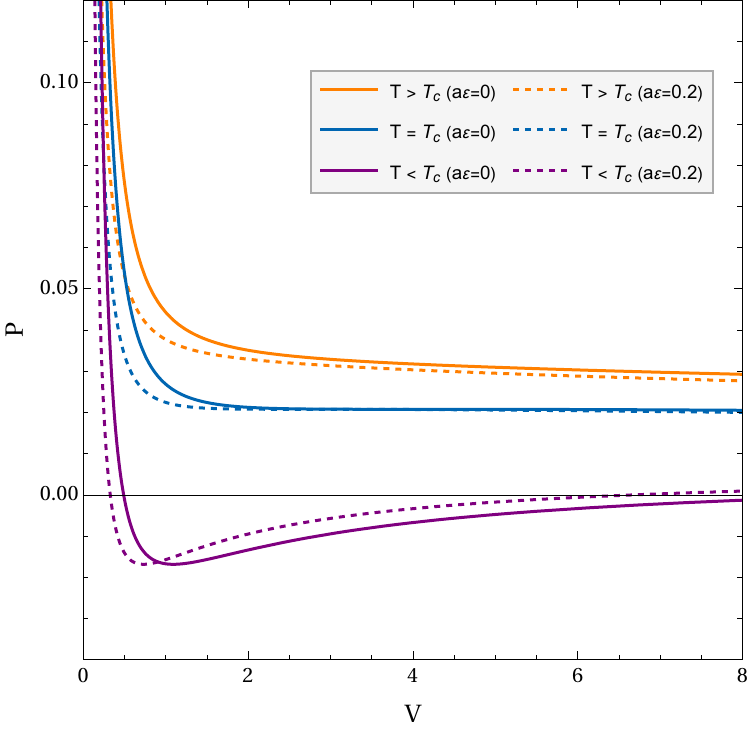}
    \caption{Variation of pressure ($P$) with (a) horizon radius ($r_{+}$) at fixed charge ($Q=1$) and (b) volume ($V$) at fixed charge ($Q=0.4$), for different $a\varepsilon$. The solid curve corresponds to the critical isotherm $T=T_c$, while dashed curves represent $T>T_c$ and $T<T_c$ as indicated.}
    \label{fig-6}
\end{figure}
The corresponding $P-r_{+}$ and $P-V$ plots are shown in Fig-\ref{fig-6}. In the $P-r_{+}$ diagram, the upper dashed line represents the `ideal gas' one-phase behavior for $T > T_{c}$. The solid line denotes the critical isotherm at $T = T_{c}$. The lower dashed line corresponds to temperatures smaller than the critical temperature. The study shows that the behavior resembles that of a Van der Waals fluid similar to that discussed in \cite{Mann}. We observe that the pressure variation with horizon radius does not depend on the NC modification. $P-V$ diagram also shows that at $T>T_{c}$, the system behaves like an ideal gas. The critical isotherm $T = T_{c}$ indicates an inflection point at the critical volume $v_{c}$ and the critical pressure $P_{c}$. At $T<T_{c}$, $P-V$ isotherms indicate an unstable thermodynamic system and a first-order phase transition between small and large black holes. This analysis clearly shows that the $P-V$ diagram resembles that of a Van der Waals fluid. Similar nature of $P-V$ criticality has been studied in \cite{Q-AdS} for quantum-corrected-AdS black hole. Further we observe that the NC parameter $a\varepsilon$ also affects the thermodynamic behavior of the system. Due to the presence of non-commutativity, the minimum value of pressure $P$ decreases at the same temperature $T$. This indicates that the $P-V$ structure remains intact, but it modifies the critical point and the coexistence region associated with the first-order phase transition.

In order to study the phase transitions of the system further, using Eq.(\ref{N17}), Eq.(\ref{N15}), and Eq.(\ref{N19}), we find the $\kappa$-deformed Gibbs free energy as 
\be 
G = H - TS = \frac{e^{-2a\varepsilon} r_+}{4}\left(1 + \frac{3Q^2}{r_+^2} - \frac{r_+^2}{L^2}\right). \label{N27}
\ee
Further using Eq.(\ref{N18}) we rewrite the above expression as
\be 
G = \frac{e^{-2a\varepsilon} r_+}{4}\left(1 + \frac{3Q^2}{r_+^2} - \frac{8 \pi r_+^2 P}{3}\right) .\label{N28}
\ee
In the limit $a \rightarrow 0$, the above equation will give an expression of Gibbs free energy calculated in \cite{Mann}.
\begin{figure*}[t!]
    \centering
    \includegraphics[width=0.45\textwidth]{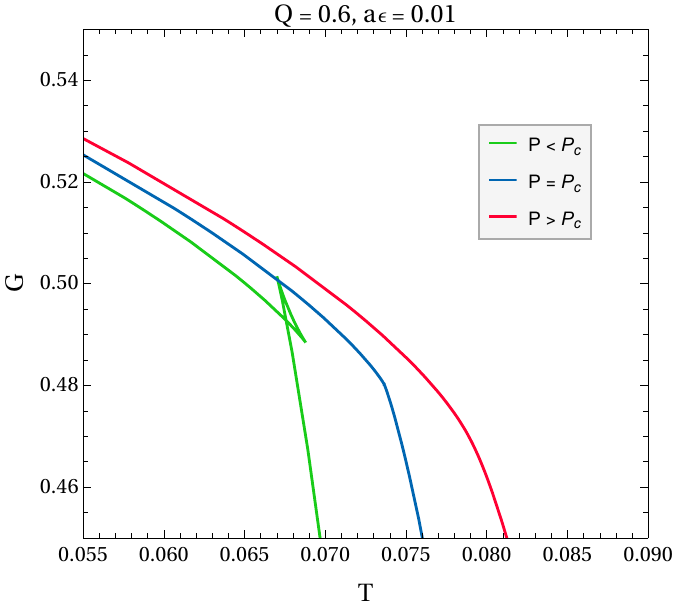}
    \includegraphics[width=0.45\textwidth]{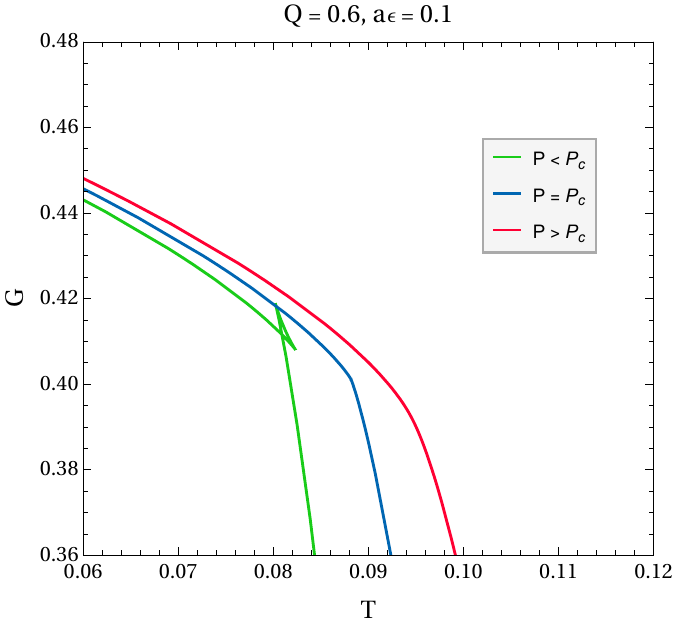}
    \includegraphics[width=0.44\textwidth]{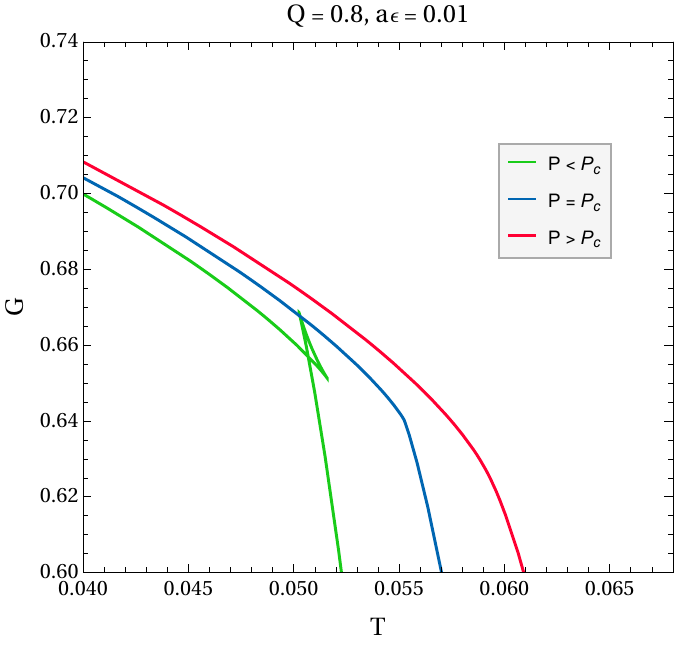}~
    \includegraphics[width=0.457\textwidth]{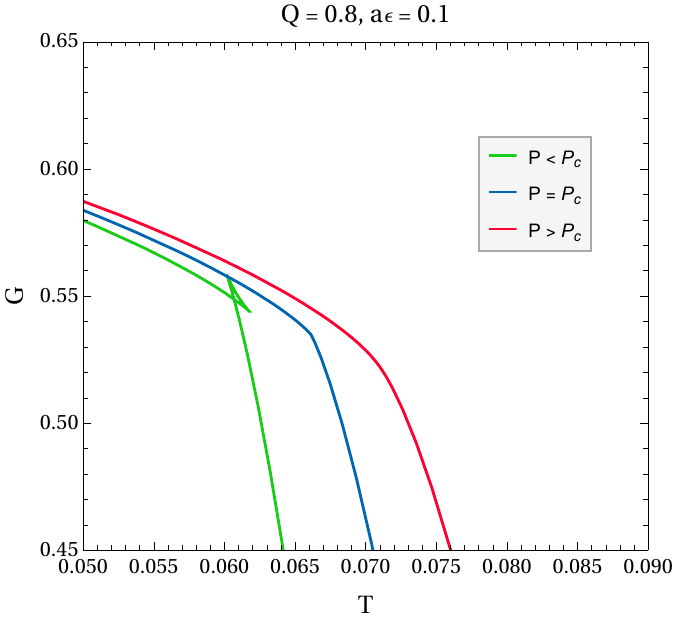}
    \caption{Variation of Gibbs free energy ($G$) as a function of temperature ($T$) for different values of $Q$ and $a\varepsilon$. The curves show the characteristic ``swallowtail'' structure for $P<P_c$, indicating a first-order phase transition; at $P=P_c$ the transition becomes second-order; for $P>P_c$ no transition occurs.}
    \label{fig-7}
\end{figure*}
To understand the thermodynamic behavior of the $\kappa$-deformed RN AdS black hole, we investigate the Gibbs free energy as a function of temperature ($T$) in Fig-\ref{fig-7}. The $G-T$ curves show small changes with variations in $Q$ and $a\varepsilon$. Their shapes remain similar for different values of the critical pressure $P_c$. All four diagrams exhibit a transition from a smooth curve to a `swallowtail' structure \cite{Charged-AdS,Mann}. This indicates a signature of a first-order phase transition. In each diagram, the green line for $P<P_{c}$ signals a phase transition between a Small Black Hole and a Large Black Hole. The curve for $P=P_{c}$ (blue line) marks the critical point where the first-order transition becomes second-order. For $P>P_{c}$, the red curves in all four diagrams are smooth, which indicates no phase transition. This implies that even NC black holes change size continuously as temperature increases, similar to Van der Waals fluids above the critical point. Increasing charge while keeping $a\varepsilon$ fixed, the transition points move to lower temperatures. On the other hand, increasing $a\varepsilon$ with a fixed charge lowers the depth of the `swallowtail' but shifts the transition point to a higher temperature. It is important to note that changes in $G-T$ plots due to charge or the non-commutativity of space-time resemble the behavior of Van der Waals fluids, indicating analogy between $\kappa$-deformed black hole and classical fluid.

\section{Deformed Joule-Thomson expansion}\label{sec:5}

\begin{figure}[h!]
    \centering
    \includegraphics[width=0.48\textwidth]{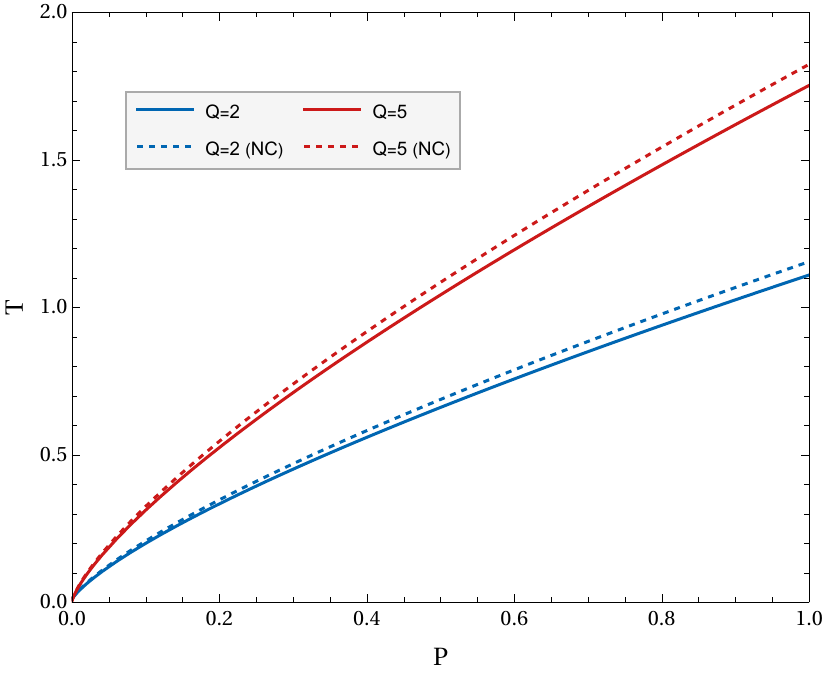}
  \caption{Inversion curves for different values of $Q$ and $a\varepsilon$. Solid lines: commutative case; dashed lines: $\kappa$-deformed case. The upward shift indicates an enlarged cooling region.}
    \label{fig-8}
\end{figure}

\begin{figure*}[t!]
    \centering
    \includegraphics[width=0.47\textwidth]{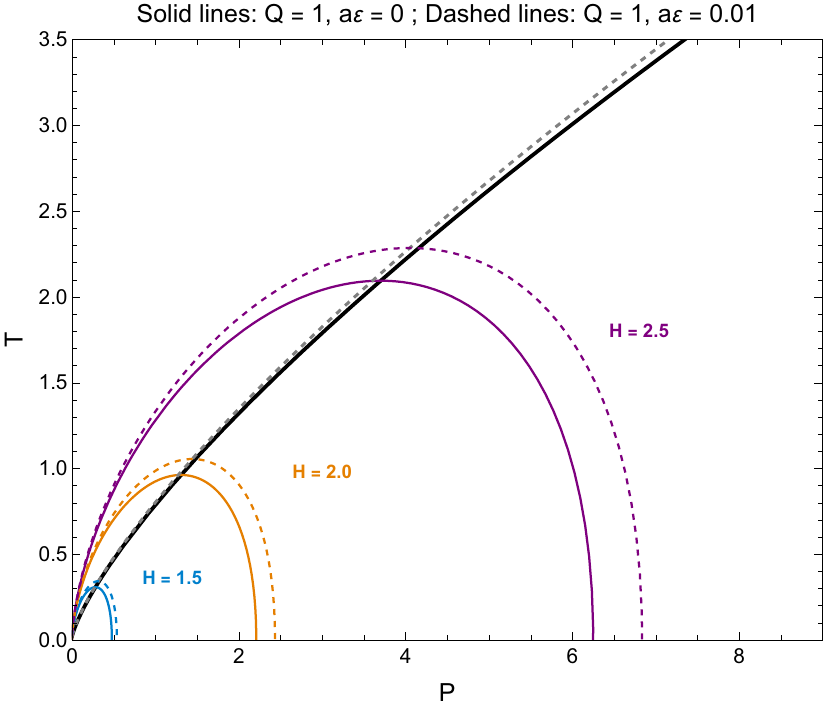}~
    \includegraphics[width=0.47\textwidth]{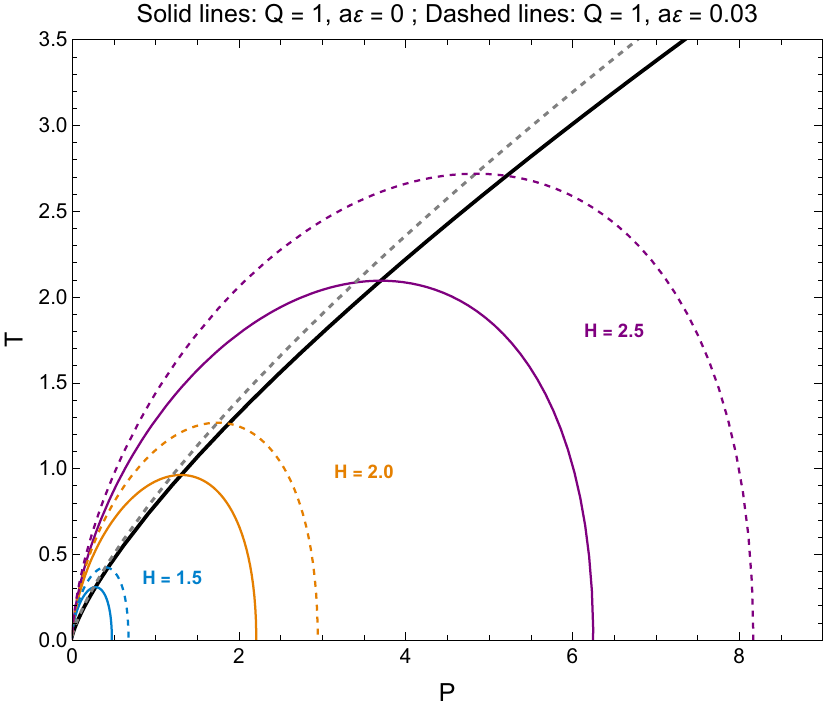}
    \includegraphics[width=0.48\textwidth]{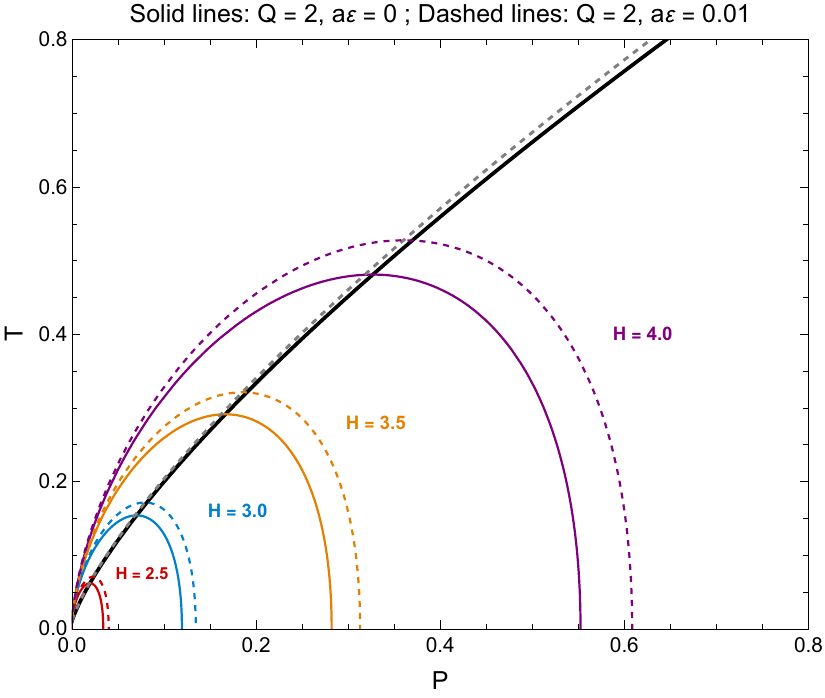}
    \includegraphics[width=0.48\textwidth]{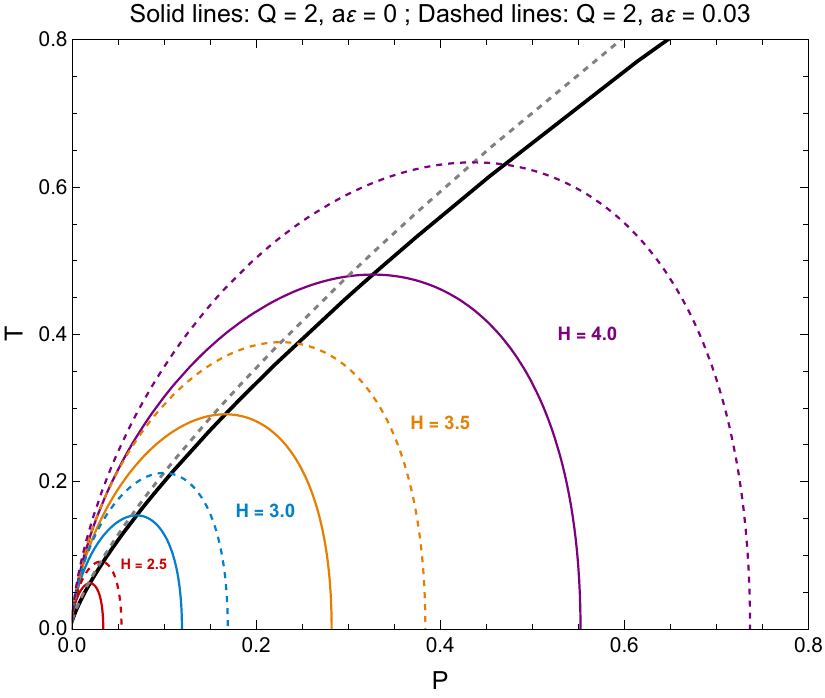}
    \caption{Inversion-isenthalpic curves for different values of $Q$ and $a\varepsilon$. Solid lines: commutative; dashed lines: NC corrected. The black lines denote the inversion curves (coexistence boundaries). Above the inversion curves, cooling occurs; below, heating occurs. NC corrections shift the isenthalpic curves upward, broadening the cooling region.}
    \label{fig-9}
\end{figure*}

In this section, we examine the effect of space-time non-commutativity on the Joule-Thomson expansion \cite{J-Thom} for the $\kappa$-deformed RN AdS black hole. The Joule-Thomson expansion is a throttling process describing the change in temperature of a thermodynamic system (such as a gas) during an adiabatic, isenthalpic expansion. In this process, the gas is first pressurized and subsequently experiences adiabatic expansion through a throttle into the low-pressure section of a thermally insulated pipe, resulting in a temperature change. The generic expression for the Joule-Thomson coefficient is given by \cite{C-AdS}
\be
\mu_{JT} = \frac{1}{C_P}\left[T\left(\frac{\partial V}{\partial T}\right)_P - V\right]. \label{N29}
\ee
Utilizing Eq.(\ref{N17}), Eq.(\ref{N18}), and Eq.(\ref{N20a}), we find
\be
T\left(\frac{\partial V}{\partial T}\right)_P - V =e^{-2a\varepsilon} 4\pi r_+^3 \left( \frac{4 + 16\pi G P r_+^2 - \frac{6Q^2}{r_+^2}}{3\left(8\pi G P r_+^2 - 1 + \frac{3Q^2}{r_+^2}\right)} \right) .\label{N30}
\ee
Now using Eq.(\ref{N21}) and the above equation in Eq.(\ref{N29}) gives the $\kappa$-deformed Joule-Thomson coefficient to be
\be
\mu_{\mathrm{JT}} = \frac{e^{2a\varepsilon} 8 r_+}{3} \frac{1 + 4\pi G P r_+^2 - \frac{6Q^2}{4r_+^2}}{1 - \frac{Q^2}{r_+^2} + 8\pi G P r_+^2}. \label{N31}
\ee
Note that the above expression also undergoes NC modifications. One can determine whether cooling or heating will occur by examining the sign of the above equation. To find the heating and cooling regions in the $T-P$ plane, we need the inversion point. We first use Eq.(\ref{N17}) and Eq.(\ref{N18}) to find the modified temperature for the $\kappa$-deformed RN AdS black hole as
\be
T = \frac{e^{2a\varepsilon}}{4\pi r_+} \left( 1 + 8\pi G P r_+^2 - \frac{Q^2}{r_+^2} \right).\label{N31b}
\ee
Setting the Joule-Thomson expansion coefficient, denoted by $\mu_{\mathrm{JT}}$ (a measure of the temperature change during a throttling process at constant enthalpy), to zero leads to
\be
1 + 4\pi G P_i r_i^2 - \frac{3Q^2}{4r_i^2} = 0. \label{N32}
\ee
To solve for the inversion temperature $T_i$, we use the above equation in Eq.(\ref{N31b}) and obtain
\be
T_i = e^{2a\varepsilon}\left(\frac{ Q^2}{8\pi r_i^3} - \frac{1}{4\pi r_i}\right). \label{N33}
\ee
Calculating Eq.(\ref{N31b}) at the inversion point and subtracting the above equation, we find
\be
8\pi G P_i r_i^4 + 2r_i^2 - 3Q^2 = 0. \label{N34}
\ee
Solving the above quadratic equation, we find the inversion radius as
\be
r_i = \frac{1}{2\sqrt{2}} \sqrt{\frac{\sqrt{1 + 24\pi P_i Q^2} - 1}{\pi P_i}} . \label{N35}
\ee
Substituting the above expression of inversion radius in Eq.(\ref{N33}) we find the $\kappa$-deformed inversion temperature to be
\be
T_i = \frac{e^{2a\varepsilon} \sqrt{P_i } \left( 1 + 16\pi P_i Q^2 - \sqrt{1 + 24\pi  P_i Q^2} \right)}{\sqrt{2\pi} \left( \sqrt{1 + 24\pi  P_i Q^2} - 1 \right)^{3/2}}. \label{N36}
\ee
In the commutative limit, $a \rightarrow 0$, the above equation reduces to the expression of inversion temperature \cite{C-AdS}. Inversion curves for various values of the charge $Q=2 ,5$ and NC parameter $a\varepsilon = 0.02$ are depicted in Fig-\ref{fig-8}. In the figure, the solid curve represents the inversion curve for the standard charged AdS black hole, while the dashed curve represents the inversion curve incorporating NC geometry corrections. We observe that, for a fixed charge, non-commutativity of space-time increases the magnitude of the inversion temperature. As a result, $T-P$ curves go upward, indicating that the cooling region is expanded. This implies that a $\kappa$-deformed RN AdS black hole can cool during isenthalpic expansion over a broader range of temperatures and pressures than a standard RN AdS black hole. 
In the limit $P_i \rightarrow 0$, the inversion radius becomes $r_i = \sqrt{\frac{3Q^2}{2}}$. Substituting this into Eq.(\ref{N33}), the minimum value of the inversion temperature is to be
\begin{equation}
T_i^{\min} = \frac{e^{2a\varepsilon}}{6\sqrt{6}\pi Q}. \label{N37}
\end{equation}
Next, we find the ratio between the minimum value of inversion temperature and the critical temperature to be
$\frac{T_i^{\min}}{T_c} = \frac{1}{2}$. This shows that the modification factor arising from non-commutativity cancels out, preserving the universal ratio obtained for standard RN AdS black holes \cite{C-AdS}. Using Eq.(\ref{N18}) in Eq.(\ref{N19}) we get the pressure as a function of $H$ and $r_+$ as
\be
P(r_+, H) = \frac{3}{4\pi r_+^3} \left( H e^{2a\varepsilon} - \frac{r_+}{2} - \frac{Q^2}{2 r_+} \right). \label{N38}
\ee
To elucidate the nature of the phase transition, $T-P$ curves (inversion-isenthalpic) were parametrically plotted using Eq.(\ref{N31b}) and Eq.(\ref{N38}), while maintaining a constant black hole enthalpy $H$, as depicted in Fig-\ref{fig-9}.
The solid curves in the figure represent the inversion-isenthalpic curves for the standard RN AdS black hole, while the dashed lines represent the NC corrections for various enthalpy values $H$. Here, both the solid black line and the dashed black line represent inversion curves (coexistence curves) for the standard RN AdS black hole and the $\kappa$-deformed RN AdS black hole, respectively. Isenthalpic curves exhibit a positive slope above the inversion curves. As a result, cooling occurs above the inversion curves. Below the inversion curves, the slope sign reverses. Consequently, heating occurs in this region. Conversely, cooling or heating does not occur on the inversion curve, which serves as a boundary separating the two regions. At the coexistence curve phase transition between a small black hole and a large black hole occurs, indicating a first-order phase transition. An increase in charge results in an upward shift of the inversion curve (see Fig-\ref{fig-8}), indicating that the black hole can cool during isenthalpic expansion across a broader range of temperatures and pressures \cite{C-AdS}. We observe a similar, though small but significant effect with NC corrections, which also cause a slight upward shift in the inversion-isenthalpic curves for a fixed charge (see Fig-\ref{fig-9}). Consequently, the black hole can cool during isenthalpic expansion over a large range of temperatures and pressures. It is important to note that for the $\kappa$-deformed RN AdS black hole, the phase transition resembles that of the Van der Waals liquid-gas system, similar to the standard and non-commutative RN AdS black holes \cite{C-AdS,NC-BH}.

\section{Conclusion}\label{sec:6}

In this work, we have systematically investigated the influence of a minimal length scale on the thermodynamics of the RN AdS black hole within the framework of $\kappa$-deformed space-time. By constructing the deformed metric and deriving all thermodynamic quantities, we have uncovered both qualitative analogies and quantitative deviations from the standard commutative case.

Our main findings are threefold. \textbf{First}, regarding thermal properties, the NC parameter $a>0$ enhances the Hawking temperature (implying faster evaporation) and suppresses the entropy (reducing the effective number of microstates), while leaving the mass-horizon relation unchanged. The heat capacity analysis reveals that the stability transition (divergence of $C_P$) is shifted in horizon radius, but the location of the divergence is independent of $a\varepsilon$—a consequence of the overall rescaling of the metric. \textbf{Second}, in the context of $P$-$V$ criticality and phase transitions, the deformed black hole retains the Van der Waals-like small-to-large black hole phase transition, with a critical ratio slightly lower than $3/8$. The NC modification lowers the coexistence pressure and alters the swallowtail depth in the Gibbs free energy, indicating a reduction in latent heat. \textbf{Third}, the Joule–Thomson expansion shows that non-commutativity increases the inversion temperature for fixed charge, thereby expanding the cooling region in the $T$-$P$ plane. Remarkably, the universal ratio $T_i^{\min}/T_c = 1/2$ remains intact despite these deformations.

These results demonstrate that the $\kappa$-deformed RN AdS black hole serves as a viable toy model for testing quantum-gravity effects on macroscopic thermodynamic processes. Future work could explore the observational consequences of this minimal length scale. In particular, computing the shadow radius and gravitational lensing signatures for the $\kappa$-deformed RN AdS black hole could provide a pathway to constrain the deformation parameter $a$ using Event Horizon Telescope (EHT) data on M87* and Sagittarius A* \cite{observ}. Additionally, extending this analysis to rotating (Kerr-AdS) black holes in $\kappa$-modified space-time would be a natural next step.

\section*{Acknowledgements}
The authors gratefully acknowledge the support of the National Natural Science Foundation of China under Grant No.  12075059, as well as the start-up fund of USTC.

\section*{Data availability statement}

Data sharing is not applicable to this article as no datasets were generated or analysed during the current study.

\section*{Conflict of interest }

The authors declare no Conflict of interest.



\end{document}